\documentclass[11pt]{article}
\usepackage{amsfonts}
\usepackage{amssymb}
\usepackage{latexsym}
\newcommand{\eq}{\begin{equation}}
\newcommand{\en}{\end{equation}}
\newcommand{\eqn}{\begin{eqnarray}}
\newcommand{\enn}{\end{eqnarray}}

\newcommand{\beq}{\begin{equation}}
\newcommand{\eeq}{\end{equation}}
\newcommand{\tn}{\ensuremath{\tilde{n}}}
\newcommand{\ta}{\ensuremath{\tilde{a}}}
\newcommand{\tb}{\ensuremath{\tilde{b}}}
\newcommand{\tc}{\ensuremath{\tilde{c}}}
\newcommand{\tx}{\ensuremath{\tilde{x}}}
\newcommand{\ty}{\ensuremath{\tilde{y}}}
\newcommand{\tz}{\ensuremath{\tilde{z}}}
\newcommand{\ti}{\ensuremath{\tilde{I}}}
\newcommand{\tj}{\ensuremath{\tilde{J}}}
\newcommand{\tk}{\ensuremath{\tilde{K}}}

\newcommand{\np}{\ensuremath{\|\varphi\|}}

\begin{document}
\begin{titlepage}
\begin{flushright}
 PSU-TH-229\\
CERN-TH/2000-126\\
\end{flushright}
\vspace{0.2cm}
\begin{center}
\begin{LARGE}
\textbf{Gauging the Full R-Symmetry Group in Five-dimensional,
 $\mathcal{N}=2$ Yang-Mills/Einstein/Tensor Supergravity}\footnote{ Work
supported
  in part by the National
Science Foundation under Grant Number PHY-9802510.}
\end{LARGE}\\
\vspace{1.0cm}
\begin{large}
M. G\"{u}naydin$^{\dagger\ddagger}$ \footnote{murat@phys.psu.edu} and
M. Zagermann$^{\ddagger}$ \end{large}\footnote{zagerman@phys.psu.edu}  \\
\vspace{.35cm}
$^{\dagger}$ \emph{CERN, Theory Division \\
1211 Geneva 23, Switzerland} \\
\vspace{.3cm}
and \\
\vspace{.3cm}
$^{\ddagger}$ \emph{Physics Department \\
Penn State University\\
University Park, PA 16802, USA} \\
\vspace{0.5cm}
{\bf Abstract}
\end{center}
\begin{small}

We show that certain five-dimensional, $\mathcal{N}=2$ Yang-Mills/Einstein
supergravity
theories admit the gauging of the \emph{full} R-symmetry group,
$SU(2)_{R}$, of the underlying $\mathcal{N}=2$ Poincar\'{e} superalgebra.
This generalizes the previously studied Abelian gaugings of
$U(1)_{R}\subset SU(2)_{R}$, and completes the construction of the most
general vector
 and tensor field coupled five-dimensional, $\mathcal{N}=2$ supergravity
theories
with gauge interactions. The gauging of $SU(2)_{R}$ turns out to be possible
only
in special cases, and leads to a new type of scalar potential. For
a large class of these theories the potential does not have any critical
points.

\end{small}

\end{titlepage}

\renewcommand{\theequation}{\arabic{section}.\arabic{equation}}
\section{Introduction}
\setcounter{equation}{0}

Five-dimensional gauged supergravity theories have been subject to a
renewed intense interest during the last three years. They offer an important
 tool in the study of the AdS/CFT-correspondence
\cite{jm,GKP,EW98,agmoo}
and have, more recently, been discussed as a potential framework for an
embedding of the Randall/Sundrum (RS-)scenario \cite{RS1,RS2} into
string/M-theory.

Whereas the embedding of the original discontinuous RS-model \cite{RS1}
into $5d$, $\mathcal{N}=2$ gauged pure supergravity on
$\mathbb{R}^{4}\times S^{1}/Z_{2}$
 was studied in \cite{ABN,GP,FLP},
a realization in terms of a  \emph{smooth} (``thick'')
BPS domain wall solution seemed to be incompatible with a variety of scalar
potentials
 of known matter coupled $\mathcal{N}=2$ supergravity theories
\cite{KL,BC2,GL}.

Since the most general $5d$, $\mathcal{N}=2$ gauged supergravity theory has
not yet
been constructed\footnote{See note added.}, it is, however, still unclear
how general
these no-go theorems really are. A construction of the most general types of
these theories should therefore help to settle this question, and might also
be
interesting for (bulk-)matter coupled generalizations of the discontinuous
model
 of \cite{RS1,ABN,GP,FLP}. At the same time, a complete knowledge of
$\mathcal{N}=2$ gauged
  supergravity theories might also contribute to a better understanding of
various
aspects of the $\mathcal{N}=8$ theory (like e.g. the structure of its vacua)
with
possible implications for the AdS/CFT-correspondence.

Motivated by these and other applications, we have recently studied
 the possible gaugings of vector and tensor field coupled $5d$,
  $\mathcal{N}=2$ supergravity theories.  All these theories (including the
ones
involving tensor multiplets) can be derived from the \emph{ungauged}
$\mathcal{N}=2$
Maxwell/Einstein supergravity theories (MESGT's) of ref. \cite{GST1}.
These theories describe the coupling of Abelian vector multiplets to
$\mathcal{N}=2$ supergravity and have a global symmetry group of the form
$SU(2)_{R}\times G$. Here, $G$ is the subgroup of the isometry group of the
scalar field target space that extends to a symmetry group of the full
Lagrangian,
 and $SU(2)_{R}$ denotes the automorphism group (``R-symmetry group'')
of the $5d$, $\mathcal{N}=2$ Poincar\'{e} superalgebra.

In \cite{gz99} we generalized the earlier work \cite{GST2} and constructed
all possible gaugings of subgroups of $U(1)_{R}\times G$, where
 $U(1)_{R}\subset SU(2)_{R}$ denotes the Abelian subgroup of $SU(2)_{R}$.
  In particular, we also covered the case when the gauging of a subgroup of $G$
   involves the dualization of some of the vector fields of the ungauged theory
 to ``self-dual'' \cite{PTvN}
    antisymmetric
tensor fields, a mechanism that is well-known from the maximally extended
gauged
supergravities in $d=7$ \cite{PPvN7} and $d=5$ \cite{GRW0,GRW1,PPvN}
dimensions.

Thus, the only gaugings that have not yet been covered in this framework
are those
 involving gaugings of the \emph{full} R-symmetry group $SU(2)_{R}$.
It is the purpose of this paper to close this gap. This will complete
the construction of the possible gaugings of the entire vector/tensor sector
of $\mathcal{N}=2$ matter coupled supergravity theories in five dimensions.

The outline of this paper is as follows. In Section 2, we first analyze to what
extent the full R-symmetry group $SU(2)_{R}$ can be gauged within the
framework
of
vector and tensor field coupled 5D, $\mathcal{N}=2$ supergravity theories.
The corresponding Lagrangians and the supersymmetry transformation rules
are then derived via the Noether method starting from some of our earlier
results \cite{gz99}.
 Section 3, finally, is devoted to a discussion of the
resulting scalar potentials.

\section{Gauging the full R-symmetry group $SU(2)_{R}$}
\setcounter{equation}{0}

The gauging of $SU(2)_{R}$  is a little less straightforward
than gaugings of subgroups of $U(1)_{R}\times G$, as we shall now explain.

The supermultiplets we are dealing with are ($\mu, \nu, \dots$ and $m,n,\ldots$
denote curved and flat spacetime indices, respectively):
\begin{enumerate}

\item The $\mathcal{N}=2$ supergravity multiplet, containing the graviton
(f\"{u}nfbein) $e_{\mu}^{m}$, two gravitini $\Psi_{\mu}^{i}$
($i,j,\ldots =1,2$) and one vector field $A_{\mu}$

\item The $\mathcal{N}=2$ vector multiplet, comprising one vector field
$A_{\mu}$, two spin-$1/2$ fermions $\lambda^{i}$ ($i,j,\ldots 1,2$) and
one real scalar field $\varphi$

\item The $\mathcal{N}=2$ ``selfdual'' tensor multiplet consisting of
two real two-form fields $B_{\mu\nu}^{(1)}$, $B_{\mu\nu}^{(2)}$; four
spin-$1/2$ fermions $\lambda^{(1)i}$,  $\lambda^{(2)i}$ ($i,j,\ldots =1,2$)
and two real scalar fields $\varphi^{(1)}$, $\varphi^{(2)}$.

\end{enumerate}

Of all the above fields, only the gravitini and the spin-$1/2$ fermions
transform non-trivially under $SU(2)_{R}$ (they form doublets labelled by the
index $i=1,2$).  In particular, all the vector fields are \emph{singlets} under
 $SU(2)_{R}$. In order to gauge a non-Abelian symmetry group like $SU(2)_{R}$,
 however, one needs vector fields that transform in the \emph{adjoint}
representation
of the gauge group.

The only way to solve this problem is to identify $SU(2)_{R}$ with
an $SU(2)$ subgroup of the scalar manifold isometry group, $G$, and to
gauge both $SU(2)$'s simultaneously. In other words, $SU(2)_{R}$ can not
be gauged by itself, rather one has to gauge a diagonal subgroup of
 $SU(2)_{R}\times SU(2)_{G}\subset SU(2)_{R}\times G $. The most natural
starting point
for a gauging of $SU(2)_{R}$ is therefore a ``Yang-Mills/Einstein
supergravity theory''
 (see \cite{GST2,gz99,gz2} for details on this terminology)
in which a subgroup $K \supset SU(2)_{G}$ of $G$ is gauged. In order to be
as general
 as possible, we consider the case when the supersymmetric
gauging of $K\subset G$
requires the introduction of tensor fields (the case without tensor fields
can easily be
 recovered as a special case).
At this point we require the gauge group $K$ only to  have an $SU(2)$
subgroup $SU(2)_{G}\subset K$, but leave it otherwise undetermined.

We start by recalling some relevant properties of Yang-Mills/Einstein
supergravity
 theories with tensor fields (see \cite{gz99,gz2} for details).
Yang-Mills/Einstein supergravity theories with tensor fields describe
the coupling of $n$ vector multiplets and $m$ self-dual tensor multiplets
to supergravity.
 Consequently, the field content of
these theories is
\begin{equation}
\{ e_{\mu}^{m}, \Psi_{\mu}^{i}, A_{\mu}^{I}, B_{\mu\nu}^{M},
\lambda^{i\ta}, \varphi^{\tx}\}
\end{equation}
where
\begin{eqnarray*}
I,J,K\ldots &=& 0,1, \ldots n \\
M,N,P\ldots &=& 1,2, \ldots 2m \\
\ta,\tb,\tc,\ldots &=& 1,\ldots, \tn\\
\tx,\ty,\tz,\ldots &=& 1,\ldots, \tn,
\end{eqnarray*}
with $\tn=n+2m$.

Note that we have combined the `graviphoton' with the $n$ vector fields
of the $n$ vector multiplets into a single $(n+1)$-plet of vector fields
$A_{\mu}^{I}$ labelled by the index $I$. Also, the spinor and scalar fields
of the vector and tensor multiplets are combined into
$\tn$-tupels of spinor and scalar fields.
The indices $\ta, \tb, \ldots$ and $\tx, \ty, \ldots$ are the
flat and curved indices, respectively, of
the $\tn$-dimensional target  manifold, $\mathcal{M}$, of the scalar fields.
The metric, vielbein and spin connection on $\mathcal{M}$ will be
denoted by $g_{\tx\ty}$, $f_{\tx}^{\ta}$ and $\Omega_{\tx}^{\ta\tb}$,
respectively.

A  subset of the vector fields $A_{\mu}^{I}$ is used to promote
a subgroup $K$ of the isometry group of $\mathcal{M}$ to a
Yang-Mills-type gauge symmetry. Apart from these gauge fields, only the
tensor fields $B_{\mu\nu}^{M}$,
the spin-$1/2$ fields $\lambda^{i \ta}$ and the scalar fields $\varphi^{\tx}$
transform non-trivially under this gauge group $K$.

The $K$-gauge covariant derivatives of
these fields are
as follows ($\nabla$ denotes the ordinary spacetime covariant derivative,
and $g$ is
 the coupling constant of the gauge group $K$)
\begin{eqnarray}
 \mathcal{D}_{\mu}
\lambda^{i\ta}
&\equiv&  \nabla_{\mu}\lambda^{i\ta}
+gA_{\mu}^{I} L_{I}^{\ta\tb}\lambda^{i\tb}  \nonumber\\
\mathcal{D}_{\mu}
\varphi^{\tx}&\equiv&
\partial_{\mu}\varphi^{\tx}+gA_{\mu}^{I}K_{I}^{\tx}\nonumber\\
\mathcal{D}_{\mu}B_{\nu\rho}^{M}&\equiv& \nabla_{\mu} B_{\nu\rho}^{M}
+gA_{\mu}^{I}\Lambda_{IN}^{M}B_{\nu\rho}^{N}.
\end{eqnarray}
Here, $K_{I}^{\tx}$ are the Killing vector fields on $\mathcal{M}$ that
generate
 the subgroup $K$ of its isometry group. The $\varphi$-dependent
matrices $L_{I}^{\ta\tb}$ and the \emph{constant} matrices $\Lambda_{IN}^{M}$
are the
 $K$-transformation matrices
 of $\lambda^{i\ta}$ and $B_{\mu\nu}^{M}$, respectively.

Denoting the curls of $A_{\mu}^{I}$ by $F_{\mu\nu}^{I}$ and the structure
 constants of $K$ by $f_{JK}^{I}$, we combine the non-Abelian field strengths
  $\mathcal{F}_{\mu\nu}^{I}=F_{\mu\nu}^{I}
+gf_{JK}^{I}A_{\mu}^{J}A_{\nu}^{K}$ with the antisymmetric tensor fields
 $B_{\mu\nu}^{M}$ to form the tensorial quantity

\begin{displaymath}
\mathcal{H}_{\mu\nu}^{\ti}:= (\mathcal{F}_{\mu\nu}^{I},B_{\mu\nu}^{M}),
\qquad (\ti,\tj,\tk,\ldots = 0,\ldots, n+2m).
\end{displaymath}

The general Lagrangian of a Yang-Mills/Einstein supergravity theory
with tensor fields is then given by \cite{gz99}
\begin{eqnarray}\label{Lagrange}
e^{-1}\mathcal{L}&=& -\frac{1}{2}R(\omega)-\frac{1}{2}
{\bar{\Psi}}_{\mu}^{i}\Gamma^{\mu\nu\rho}\nabla_{\nu}\Psi_{\rho i}-
\frac{1}{4}{\stackrel{\scriptscriptstyle{o}}{a}}_{\ti\tj}
\mathcal{H}_{\mu\nu}^{\ti}
\mathcal{H}^{\tj\mu\nu}
\nonumber\cr
& & -\frac{1}{2}{\bar{\lambda}}^{i\ta}\left(\Gamma^{\mu}\mathcal{D}_{\mu}
\delta^{\ta\tb}+
\Omega_{\tx}^{\ta\tb}\Gamma^{\mu}\mathcal{D}_{\mu}\varphi^{\tx}\right)
\lambda_{i}^{\tb}-
\frac{1}{2}g_{\tx\ty}(\mathcal{D}_{\mu}\varphi^{\tx})(\mathcal{D}^{\mu}
\varphi^{\ty})\nonumber\cr
&& -\frac{i}{2}{\bar{\lambda}}^{i\ta}\Gamma^{\mu}\Gamma^{\nu}\Psi_{\mu i}
f_{\tx}^{\ta}\mathcal{D}_{\nu}\varphi^{\tx}+ \frac{1}{4}h_{\ti}^{\ta}
{\bar{\lambda}}^{i\ta}\Gamma^{\mu}\Gamma^{\lambda\rho}\Psi_{\mu i}
\mathcal{H}_{\lambda\rho}^{\ti}
\nonumber\cr
&&+\frac{i}{2\sqrt{6}}\left(\frac{1}{4}\delta_{\ta\tb}h_{\ti}+T_{\ta\tb\tc}
h_{\ti}^{\tc}\right)
{\bar{\lambda}}^{i\ta}\Gamma^{\mu\nu}\lambda_{i}^{\tb}
\mathcal{H}_{\mu\nu}^{\ti}\nonumber\cr
&& -\frac{3i}{8\sqrt{6}}h_{\ti}\left[{\bar{\Psi}}_{\mu}^{i}
\Gamma^{\mu\nu\rho\sigma}
\Psi_{\nu i}\mathcal{H}_{\rho\sigma}^{\ti}+2{\bar{\Psi}}^{\mu i}
\Psi_{i}^{\nu}\mathcal{H}_{\mu\nu}^{\ti}\right]\nonumber\cr
&& + \frac{e^{-1}}{6\sqrt{6}}C_{IJK}\varepsilon^{\mu\nu\rho\sigma\lambda}
\left\{ F_{\mu\nu}^{I}F_{\rho\sigma}^{J}A_{\lambda}^{K} + \frac{3}{2}g
F_{\mu\nu}^{I}A_{\rho}^{J}(f_{LF}^{K}A_{\sigma}^{L}A_{\lambda}^{F})\right.
\nonumber\cr
&& \qquad\qquad\qquad\qquad+\left.
\frac{3}{5}g^{2}(f_{GH}^{J}A_{\nu}^{G}A_{\rho}^{H})
(f_{LF}^{K}A_{\sigma}^{L}A_{\lambda}^{F})A_{\mu}^{I}\right\}\nonumber\cr
&&+\frac{e^{-1}}{4g}\varepsilon^{\mu\nu\rho\sigma\lambda}\Omega_{MN}
B_{\mu\nu}^{M}\mathcal{D}_{\rho}B_{\sigma\lambda}^{N}\nonumber\\
&&+g{\bar{\lambda}}^{i\ta}\Gamma^{\mu}\Psi_{\mu i}W_{\ta}+
g{\bar{\lambda}}^{i\ta}\lambda_{i}^{\tb}W_{\ta\tb}-g^{2}P
\end{eqnarray}
with $e \equiv \det(e_{\mu}^{m})$.
The transformation laws are (to leading order in fermion fields)
\begin{eqnarray}\label{trafo}
\delta e_{\mu}^{m}&=& \frac{1}{2}{\bar{\varepsilon}}^{i}
\Gamma^{m}\Psi_{\mu i}\nonumber\cr
\delta \Psi_{\mu}^{i} &=&\nabla_{\mu}\varepsilon^{i}+\frac{i
}
{4\sqrt{6}}h_{\ti}(\Gamma_{\mu}^{\:\:\:\nu\rho}-4\delta_{\mu}^{\nu}
\Gamma^{\rho})\mathcal{H}_{\nu\rho}^{\ti}\varepsilon^{i}\nonumber\cr
\delta A_{\mu}^{I}&=& \vartheta_{\mu}^{I}\nonumber\cr
\delta B_{\mu\nu}^{M} &=& 2\mathcal{D}_{[\mu}\vartheta_{\nu]}^{M} +
\frac{\sqrt{6}g}{4}
\Omega^{MN}h_{N}{\bar{\Psi}}^{i}_{[\mu}\Gamma_{\nu]}\varepsilon_{i}
+\frac{ig}{4}\Omega^{MN}h_{N\ta}{\bar{\lambda}}^{i\ta}\Gamma_{\mu\nu}
\varepsilon_{i}\nonumber\\
\delta \lambda^{i\ta}  &=& -\frac{i}{2}f_{\tx}^{\ta}
\Gamma^{\mu}(\mathcal{D}_{\mu}
\varphi^{\tx})\varepsilon^{i} + \frac{1}{4}h_{\ti}^{\ta}
\Gamma^{\mu\nu}
\varepsilon^{i}\mathcal{H}_{\mu\nu}^{\ti}+gW^{\ta}\varepsilon^{i}\nonumber \\
\delta \varphi^{\tx}&=&\frac{i}{2}f^{\tx}_{\ta}{\bar{\varepsilon}}^{i}
\lambda_{i}^{\ta}
\end{eqnarray}
with
\begin{equation}
\vartheta_{\mu}^{\ti}\equiv -\frac{1}{2}h_{\ta}^{\ti}{\bar{\varepsilon}}^{i}
\Gamma_{\mu}\lambda_{i}^{\ta}+\frac{i\sqrt{6}}{4}h^{\ti}
{\bar{\Psi}}_{\mu}^{i}\varepsilon_{i}.
\end{equation}

The various scalar field dependent quantities
 $\stackrel{\scriptscriptstyle{o}}{a}_{\ti\tj}$, $h_{\ti}$, $h^{\ti}$,
$h_{\ti}^{\ta}$,
  $h^{\ti\ta}$ and
$T_{\ta\tb\tc}$ that contract the different types of indices are already
present in
 the corresponding \emph{ungauged} MESGT's and describe the ``very special''
geometry
 of the scalar manifold $\mathcal{M}$ (see \cite{GST1} for details).
These ungauged MESGT's also contain a constant symmetric tensor
$C_{\ti\tj\tk}$.
If the gauging of $K$ involves the introduction of tensor fields, the
coefficients of
 the type $C_{MNP}$ and $C_{IJM}$ have to vanish \cite{gz99}. The only
components that survive such a  gauging are thus $C_{IJK}$, which appear in
the
Chern-Simons-like  term of (\ref{Lagrange}), and $C_{IMN}$, which
are related to  the transformation matrices of the tensor fields by
\begin{displaymath}
\Lambda_{IN}^{M}=\frac{2}{\sqrt{6}}\Omega^{MP}C_{IPN}.
\end{displaymath}
Here $\Omega^{MN}$ is the inverse of  $\Omega_{MN}$, which is a (constant)
invariant
antisymmetric tensor of the gauge
group $K$:
\begin{equation}
\Omega_{MN}=-\Omega_{NM}, \qquad \Omega_{MN}\Omega^{NP}=\delta_{M}^{P}.
\end{equation}

The quantities $W^{\ta}(\varphi)$ and $W^{\ta\tb}(\varphi)$ and the scalar
 potential $P(\varphi)$ are due to the gauging of $K$ in the presence of the
tensor
  fields, and are given by
\begin{eqnarray}
W^{\ta}&=&-\frac{\sqrt{6}}{8}h_{M}^{\ta}\Omega^{MN}h_{N}\nonumber\\
W^{\ta\tb}&=&-W^{\tb\ta}= ih_{\, }^{J[\ta}K_{J}^{\tb]}+\frac{i\sqrt{6}}{4}
h^{J}K_{J}^{\ta;\tb}\nonumber\\
P&=&2W^{\ta}W^{\ta},
\end{eqnarray}
where the semicolon denotes covariant differentiation on the target space
$\mathcal{M}$.

We will now use the above  theory as our starting point for the additional
gauging
 of $SU(2)_{R}$. To this end, we first split the index $I$ of the $(n+1)$
vector fields
  $A_{\mu}^{I}$
according to
\begin{displaymath}
I=(A,I'),
\end{displaymath}
where $A, B, C, \ldots\in \{1,2,3\}$ are the indices corresponding
to the three gauge fields of $SU(2)_{G}\subset K$, and $I',J',K', \ldots$
label the remaining $(n-2)$  vector fields.

In order to gauge $SU(2)_{R}$, we use the gauge fields $A_{\mu}^{A}$
to covariantize the $K$- and spacetime covariant derivatives of the fermions
also with respect to $SU(2)_{R}$, i.e., we make the replacements
\begin{eqnarray}
\nabla_{\mu}\Psi_{\nu}^{i}&\longrightarrow& \mathfrak{D}_{\mu}\Psi_{\nu}^{i}
:=\nabla_{\mu}\Psi_{\nu}^{i} +
g_{R} A_{\mu}^{A} \Sigma_{A j}^{\,\,\,\,i} \Psi_{\nu}^{j}\\
\nabla_{\mu}\varepsilon^{i}&\longrightarrow& \mathfrak{D}_{\mu}\varepsilon^{i}
\,\,\,:=\nabla_{\mu}\varepsilon^{i} +
g_{R} A_{\mu}^{A} \Sigma_{A j}^{\,\,\,\,i} \varepsilon^{j}\\
\mathcal{D}_{\mu} \lambda^{i\ta}&\longrightarrow &\mathfrak{D}_{\mu}
 \lambda^{i\ta}:=\mathcal{D}_{\mu}\lambda^{i\ta}
 + g_{R} A_{\mu}^{A} \Sigma_{A j}^{\,\,\,\,i} \lambda^{j\ta}\nonumber\\
&&\qquad\quad\,\equiv
\nabla_{\mu}\lambda^{i\ta}
+gA_{\mu}^{I} L_{I}^{\ta\tb}\lambda^{i\tb}
+ g_{R} A_{\mu}^{A} \Sigma_{A j}^{\,\,\,\,i} \lambda^{j\ta}
\end{eqnarray}
in the Lagrangian (\ref{Lagrange}) and the transformation laws (\ref{trafo}).
Here, $g_{R}$ denotes the $SU(2)_{R}$ coupling constant, and the
$\Sigma_{A j}^{\,\,\,\,i}$  ($i,j, \ldots =1,2$)
are the $SU(2)_{R}$ transformation matrices of the fermions, which can be
chosen as
\begin{equation}\label{Pauli}
\Sigma_{Aj}^{\,\,\,\,i}=\frac{i}{2} \sigma_{Aj}^{\,\,\,\,i}
\end{equation}
with $\sigma_{Aj}^{\,\,\,\,i}$ being the Pauli matrices. The indices
$i,j,\ldots$
are raised and lowered according to
\begin{displaymath}
  X^i=\varepsilon ^{ij}X_j\,,\qquad X_i=X^j\varepsilon _{ji}\,
\end{displaymath}
with $\varepsilon^{ij}$, $\varepsilon_{ij}$ antisymmetric and
$\varepsilon^{12}=\varepsilon_{12}=1$.
(The tracelessness of $\Sigma_{Aj}^{\,\,\,\,i}$ then
 implies $\Sigma_{Aij}=\Sigma_{Aji}$.)

The above replacements break supersymmetry, but the latter can be
restored by adding
\begin{eqnarray}
e^{-1}\mathcal{L}' &=& g_{R} {\bar{\Psi}}_{\mu}^{i}
\Gamma^{\mu\nu}
\Psi_{\nu}^{j} R_{0ij}(\varphi) +g_{R}{\bar{\lambda}}^{i\ta}
\Gamma^{\mu}
\Psi_{\mu}^{j}R_{\ta ij}(\varphi)\nonumber\\
&+& g_{R}{\bar{\lambda}}^{i\ta}
\lambda^{j\tb}R_{\ta\tb ij}(\varphi)-g_{R}^{2}P^{(R)}
(\varphi),
\end{eqnarray}
to the  Lagrangian and by adding
\begin{eqnarray}
\delta' \Psi_{\mu i}&=&\frac{2}{3}g_{R}R_{0ij}(\varphi)\Gamma_{\mu}
\varepsilon^{j}\nonumber\\
\delta' \lambda_{i}^{\ta}&=& g_{R}R^{\ta}_{ij}(\varphi)\varepsilon^{j}
\end{eqnarray}
to the transformation laws.

The quantities $R_{0ij}$, $R^{\ta}_{ij}$, $R_{\ta\tb ij}$ and the
additional potential term  $P^{(R)}$ are fixed by supersymmetry:
\begin{eqnarray}
R_{0ij}&=&i \sqrt{\frac{3}{8}} h^{A}\Sigma_{Aij}\\
R^{\ta}_{ij}&=& h^{A\ta}\Sigma_{Aij}\\
R_{\ta\tb ij}&=& -\frac{1}{3} \delta_{\ta\tb} R_{0ij} -i\sqrt{\frac{2}{3}}
T_{\ta\tb\tc}R^{\tc}_{ij}\\
P^{(R)}&=&-\frac{16}{3} R_{0j}^{\,\,i}R_{0\,i}^{\,\,j}-
R^{\ta i}_{\,\,\,\,j}R^{\ta j}_{\,\,\,\,\,i}.\label{PR}
\end{eqnarray}
Supersymmetry also requires
\begin{eqnarray}
f_{I'B}^{A}=f_{I'J'}^{A}&=&0\label{structure}\\
g_R{\left[ \Sigma_{A},\Sigma_{B} \right]}_{ij}&=& g f_{AB}^{C}
\Sigma_{Cij} \label{grg} \\
\Sigma_{Aij,\tx}&=&0.
\end{eqnarray}
(The structure constants of the type $f_{I'A}^{J'}$ do not necessarily
have to vanish
for supersymmetry. If they do vanish, however,  $K$ is a direct product
of $SU(2)_{G}$
and some other group $K'$. Otherwise, $K$ is a semi-direct product of the form
$(SU(2)_{G} \times S )\ltimes T $,
where $\ltimes$ denotes the semi-direct product
 and $S$ and $T$ are some other subgroups of $K$.)

The following constraints are consequences of the above and are needed
in the proof of supersymmetry:
\begin{eqnarray}
R^{\ta}_{ij} K_{J}^{\ta}&=&-\sqrt{\frac{3}{2}}\Sigma_{Aij}f^{A}_{JK} h^{K}\\
-i\sqrt{\frac{3}{2}}h^{B\ta}f_{BC}^{A}h^{C}\Sigma_{Aij}&=&
\frac{10}{3}W^{\ta}R_{0ij}+2R^{\tb}_{ij}W_{\ta\tb}+2R_{\ta\tb ij} W^{\tb}\\
R^{\ta}_{ij;\tx}&=&if_{\tx}^{\ta}R_{0ij}-iR_{\ta\tb ij} f_{\tx}^{\tb}\\
R_{0ij,\tx}&=&-\frac{i}{2}R_{\tx ij}.
\end{eqnarray}
Furthermore, the cancellation of  the $\delta \varphi^{\tx}$ variation
of $P^{(R)}$ and similar
terms requires that
\begin{equation}
\textrm{tr}(\Sigma_{(A}\Sigma_{B)}\Sigma_{C})=0
\end{equation}
(as well as $\textrm{tr}(\Sigma_{A})=0$), which is, however,  a general
property
of traceless antihermitian $(2\times 2)$- matrices.

For the sake of concreteness, let us conclude this section with
 a brief overview of the most
interesting examples of Yang-Mills/Einstein supergravity theories that admit
 the gauging of $SU(2)_{R}$.

Even though the constraints from supersymmetry (eq. (\ref{structure}))
allow  $K$  to be a semi-direct product
group, we shall restrict ourselves to  gauge groups that are not of the
semi-direct type.
 In this case, $K$ is
 a \emph{direct} product of $SU(2)_G$ with another group.
We can thus confine ourselves to the case $K=SU(2)_{G}$, since
additional group factors in $K$ do not change the structure of the above
theory very much.

Now, to be able to gauge  $K=SU(2)_{G}$, the isometry group of $\mathcal{M}$
must have an $SU(2)$ subgroup, $SU(2)_{G}$, that extends to a symmetry group
of the full Lagrangian with three of the vector fields of the theory
transforming   in
the adjoint representation of $SU(2)_{G}$.

For the generic Jordan
family of MESGT's with the scalar manifold $SO(1,1)\times
SO(\tn-1,1)/SO(\tn-1)$ \cite{GST1}, such a subgroup exists for all
theories with $\tn>3$ (see Section 3.1). Similarly, for the generic
non-Jordan family with the scalar manifold $SO(\tn,1)/SO(\tn)$ \cite{GST3}
one can gauge
$SU(2)_G$ whenever
$\tn>3$ (see Section 3.3).

Of the `magical' $\mathcal{N}=2$ MESGT's \cite{GST1},
all but the the one defined by the Jordan algebra of real symmetric
$(3\times 3)$-matrices, $J_3^\mathbb{R} $, admit such a gauging.

Finally, all the members of the infinite family with $SU(N)$ isometries $(N>3)$
described in ref. \cite{gz99} also admit a gauging of $SU(2)_{G}$ (and thus
of $SU(2)_{R}$).

For the generic Jordan  and the generic non-Jordan families one can choose
 the $SU(2)_{G}$ subgroup of the  isometry group such that all the
 other vector fields are inert under it, i.e. one does not have to dualize
any vector fields to
tensor fields. On the other hand,  the gauging of $SU(2)_{G}$ requires the
dualization of some of the vector fields to tensor fields in the magical
 theories as well as in the theories with $SU(N)$ isometries.

\section{The scalar potential}
\setcounter{equation}{0}

As seen in the previous section, the gauging of $SU(2)_{R}$
introduces an additional contribution, $P^{(R)}(\varphi)$,  to the total
scalar potential. Before we take a closer look at this potential,
let us first make contact with the earlier work \cite{GST2}
on the most general gauging of a $U(1)_{R}$ subgroup of $ SU(2)_{R}$.
We first note that the triplet of vector fields transforming in the adjoint
representation of $SU(2)_{G}$ cannot include the graviphoton. This follows
 from the fact that $SU(2)_{G}$ is a subgroup of the compact part
of the isometry group of $\mathcal{M}$, under which the graviphoton is inert.
 Thus the $U(1)_{R}$ gauged theories obtained by restricting oneself to a
$U(1)$ subgroup of $SU(2)_{R}$ do not describe the most general $U(1)_{R}$ gaugings possible :
For the most general $U(1)_{R}$ gauging, one can choose an arbitrary
linear combination $A_{\mu}^{I}V_{I}$ of all the vector fields, as was done
in \cite{GST2}, \emph{including}  the graviphoton.

For the theories of the Jordan family, it was shown in \cite{GST2} that the
generic $U(1)_{R}$
gauging either leads to a flat potential with Minkowski ground states
whenever
$V_{I}$ corresponds to an idempotent of the Jordan algebra, or an
Anti-de Sitter ground state whenever $V_{I}$ lies in
the ``domain of positivity'' of the
Jordan algebra, or to no critical points at all when none of the above is
true for $V_{I}$.
Looking now at the $U(1)_{R}$ restrictions of the $SU(2)_{R}$
gaugings in the Jordan family, one finds that the $V_{I}$ are of the
last type, i.e., they are
neither idempotents nor do they lie in the domain of positivity. This
already suggests that,
at least in the Jordan family, the $SU(2)_{R}$ gauging leads to theories
without critical points.

In fact, we are able to verify this statement for all theories for
which the scalar manifold $\mathcal{M}$ is a symmetric space. These can be
divided into three families:

\begin{enumerate}
\item The generic Jordan family
 \item The magical Jordan family
\item The generic non-Jordan family
\end{enumerate}
Before we look    at each of these three
families in more detail,   let us recast the scalar potential $P^{(R)}$
of eq. (\ref{PR}) into a more compact form. In the basis (\ref{Pauli}),
$P^{(R)}$ becomes

\begin{equation}\label{PR2}
P^{(R)}= \left[ -h^{A}h^{B}+\frac{1}{2}h^{A\ta}h^{B\ta} \right] \delta_{AB}.
\end{equation}
Using \cite{GST1}
\begin{displaymath}
C_{\ti\tj\tk}h^{\tk}=h_{\ti}h_{\tj}-\frac{1}{2}h_{\ti}^{\ta}h_{\tj}^{\ta},
 \end{displaymath}
this can  be rewritten as
\begin{equation}\label{PR3}
P^{(R)}=-C^{AB\ti}h_{\ti} \delta_{AB},
\end{equation}
where we have defined
\begin{displaymath}
C^{\ti\tj\tk}\equiv{\stackrel{\scriptscriptstyle{\circ}}{a}}^{\ti\ti'}
{\stackrel{\scriptscriptstyle{\circ}}{a}}^{\tj\tj'}
{\stackrel{\scriptscriptstyle{\circ}}{a}}^{\tk\tk'}C_{\ti'\tj'\tk'}
\end{displaymath}
with ${\stackrel{\scriptscriptstyle{\circ}}{a}}^{\ti\tj}$ being the inverse
of ${\stackrel{\scriptscriptstyle{\circ}}{a}}_{\ti\tj}$.

Let us now analyse this potential for the above-mentioned three families of
symmetric
spaces.

\subsection{The generic Jordan family}
The generic Jordan family corresponds to the scalar manifolds of the form
$\mathcal{M}=SO(1,1)\times SO(\tilde{n}-1,1)/SO(\tilde{n}-1)$. The latter
can be
described as the hypersurface $N(\xi)=1$ of the cubic polynomial \cite{GST1}
\begin{eqnarray}
N(\xi)&=&\left(\frac{2}{3}\right)^{\frac{3}{2}}C_{\ti\tj\tk}
\xi^{\ti}\xi^{\tj}\xi^{\tk}\nonumber\\
&=& \sqrt{2} \xi^{0}\left[ (\xi^{1})^{2} -(\xi^{2})^{2}-\ldots-
(\xi^{\tn})^{2}\right],
\end{eqnarray}
where  the $\xi^{\ti}$ parametrize an ambient space $\mathbb{R}^{\tn+1}$.
The isometry group of this space is $SO(1,1)\times SO(\tilde{n}-1,1)$. For
$SU(2)\sim SO(3)$ to be  a subgroup, one obviously needs $\tn \geq 4$, as
we will assume from now on.

The constraint $N(\xi)=1$ can be solved by
\begin{eqnarray}
\xi^{0}&=&\frac{1}{\sqrt{2} \np^{2}}\nonumber\\
\xi^{1}&=&\varphi^{1}\nonumber\\
&\vdots&\nonumber\\
\xi^{\tn}&=&\varphi^{\tn},
\end{eqnarray}
where $\np^{2}\equiv (\varphi^{1})^{2}-(\varphi^{2})^{2}-\ldots -
(\varphi^{\tn})^{2}$ has been introduced. As explained in \cite{gz2},
the scalar field metric $g_{\tx\ty}$ and the vector field metric
${\stackrel{\scriptscriptstyle{\circ}}{a}}_{\ti\tj}$ are positive
definite only for $\np^{2}>0$.
Without loss of generality, we choose $A_{\mu}^{2}$, $A_{\mu}^{3}$,
$A_{\mu}^{4}$ as the $SO(3)$ gauge fields.

For the Jordan cases, one has $C_{\ti\tj\tk}=C^{\ti\tj\tk}=\textrm{const.}$
(componentwise) \cite{GST1}. Using $h_{\ti}=\frac{1}{\sqrt{6}}
\frac{\partial}{\partial \xi^{\ti}}N|_{N=1}$ \cite{GST1}, one then
obtains for the scalar potential (\ref{PR3})
\begin{equation}
P^{(R)}=\frac{3}{2}\np^{2}.
\end{equation}
It is easy to see that this scalar potential does \emph{not} have any
critical points in the physically relevant region $\np^{2}>0$.

This situation does not change when one gauges an  additional
$SO(2)\subset G$ along
 the lines of ref.  \cite{gz2} by introducing  tensor fields.
For such a gauging, one needs at least $\tn \geq 6$. Choosing
$\xi^{5}$ and $\xi^{6}$ to form an $SO(2)$ doublet, the corresponding
vector fields
 $A_{\mu}^{5}$ and $A_{\mu}^{6}$ have to be dualized to tensor fields.
This gives
  rise to the additional potential term \cite{gz2}
\begin{equation}
P= \frac{1}{8} \frac{\left[(\varphi^{5})^{2}+
(\varphi^{6})^{2}\right]}{\np^{6}}.
\end{equation}
It is easy to verify that the combined potential $P_{tot}=P^{(R)}+P$
does not have any ground states either.

\subsection{The magical Jordan family}
We now turn to the magical Jordan family \cite{GST1}. The simplest
example in which $SU(2)_{R}$ can be gauged is provided by the
model with the scalar manifold $\mathcal{M}=SL(3,\mathbb{C})/SU(3)$.
This theory contains eight vector multiplets (i.e. it comprises
eight scalar fields and
nine vector fields). $\mathcal{M}$ can be described as the hypersurface
$N(\xi)=1$ of the cubic polynomial
\begin{equation}
N(\xi)=\sqrt{2}\xi^{4}\eta_{\alpha\beta}\xi^{\alpha}\xi^{\beta}+\gamma_{\alpha
MN}\xi^{\alpha}\xi^{M}\xi^{N},
\end{equation}
where
\begin{eqnarray}
\alpha, \beta,\ldots &=&0,1,2,3\nonumber\\
M,N,\ldots &=& 5,6,7,8\nonumber\\
\eta_{\alpha\beta}&=& \textrm{diag}(+,-,-,-)\nonumber\\
\gamma_{0}&=&-\mathbf{1}_{4}\nonumber\\
\gamma_{1}&=&\mathbf{1}_{2}\otimes \sigma_{1}\nonumber\\
\gamma_{2}&=&-\sigma_{2}\otimes\sigma_{2}\nonumber\\
\gamma_{3}&=&\mathbf{1}_{2}\otimes \sigma_{3}\nonumber.
\end{eqnarray}
It is easy to show that the vector field metric
${\stackrel{\scriptscriptstyle{o}}{a}}_{\ti\tj}$
becomes degenerate, when $\eta_{\alpha\beta}\xi^{\alpha}\xi^{\beta}=0$.
We therefore can restrict ourselves to the region
$\eta_{\alpha\beta}\xi^{\alpha}\xi^{\beta}\neq 0$, where the constraint
$N(\xi)=1$ can be solved by
\begin{eqnarray}
\xi^{\alpha}&=&\varphi^{\alpha}=:x^{\alpha}\nonumber\\
\xi^{4}&=&\frac{1-b^{T}\bar{x} b}{\sqrt{2}\|x\|^{2}}\nonumber\\
\xi^{M}&=&\varphi^{M}=:b^{M}\nonumber,
\end{eqnarray}
where $b^{T}\bar{x}b\equiv b^{M}\bar{x}_{MN}b^{N}$ with
$\bar{x}_{MN}\equiv x^{\alpha}\gamma_{\alpha MN}$ and $\|x\|^{2}
\equiv \eta_{\alpha\beta}x^{\alpha}x^{\beta}$.

In the above model, one can gauge a $(U(1)\times SU(2))$-subgroup of the
isometry group $SL(3,\mathbb{C})$. The vector field $A_{\mu}^{0}$ corresponds
to the
$U(1)$ gauge field, whereas the vector fields $A_{\mu}^{1}$, $A_{\mu}^{2}$,
$A_{\mu}^{3}$ act as the $SU(2)$ gauge fields. The vector fields $A_{\mu}^{M}$
are charged under $(U(1)\times SU(2))$ and have to be dualized to tensor
fields. The vector field $A_{\mu}^{4}$ is a spectator vector field. The
introduction of the tensor fields leads to a non-trivial potential $P$,
which turns out to be
\begin{equation}
P=-\frac{1}{8}b^{T}(\bar{x})^{3}b.
\end{equation}
As described earlier, the $SU(2)_{G}$ gauge fields $A_{\mu}^{1}$,
$A_{\mu}^{2}$,
$A_{\mu}^{3}$ can be used to simultaneously gauge $SU(2)_{R}$.
This leads to an additional potential
\begin{equation}
P^{(R)}=\frac{3}{2}\|x\|^{2}.
\end{equation}
Taking into account that $\det((\bar{x})^{3})=[\|x\|^{2}]^{6}$, it easy
to verify that the total potential $P_{tot}=P+P^{(R)}$ does \emph{not}
have any critical points in the physically relevant region, where
$\|x\|^{2}\neq 0$.

The other magical theories corresponding to $\mathcal{M}=SU^*(6)/USp(6)$ and
$\mathcal{M}=E_{6(-26)}/F_{4}$, which also allow the gauging of $SU(2)_{R}$,
have a very similar structure to the above and contain the
$SL(3,\mathbb{C})/SU(3)$ model as a subsector; one therefore does not expect
to find any
critical points either.

\subsection{The generic non-Jordan family}
This leaves us with the theories of the generic non-Jordan family \cite{GST3}.
They are given by $\mathcal{M}=SO(1,\tn)/SO(\tn)$, which can be described
as the hypersurface $N(\xi)=1$ of
\begin{equation}
N(\xi)=\sqrt{2} \xi^{0}(\xi^{1})^{2}- \xi^{1}\left[ (\xi^{2})^{2}+\ldots+
(\xi^{\tn})^{2}\right]
\end{equation}

The constraint $N=1$ can be solved by
\begin{eqnarray}
\xi^{0}&=&\frac{1}{\sqrt{2}(\varphi^{1})^{2}} +
\frac{1}{\sqrt{2}}\varphi^{1}\left[(\varphi^{2})^{2}+
\ldots +(\varphi^{\tn})^{2}\right]\nonumber\\
\xi^{1}&=&\varphi^{1}\nonumber\\
\xi^{2}&=&\varphi^{1}\varphi^{2}\nonumber\\
&\vdots&\nonumber\\
\xi^{\tn}&=&\varphi^{1}\varphi^{\tn}\nonumber
\end{eqnarray}

In contrast to the Jordan families, one no longer has the equality of
the constant
 $C_{\ti\tj\tk}$ to  $C^{\ti\tj\tk}$.  Instead, the $C^{\ti\tj\tk}$ are now
 scalar field dependent, which makes a similar analysis more
 complicated.
 What makes the calculation of the scalar potential nevertheless feasible
is that
the scalar field metric $g_{\tx\ty}$ becomes diagonal, and therefore easily
invertible, in the above coordinate system. To be specific, one obtains
\begin{equation}
g_{\tx\ty}=\textrm{diag}[3/(\varphi^{1})^{2}, (\varphi^{1})^{3}, \ldots,
(\varphi^{1})^{3}],
\end{equation}
which is positive definite for $\varphi^{1}>0$.

In order to gauge an  $SO(3)\sim SU(2)$ subgroup of the isometry group of
$\mathcal{M}$, one obviously needs at least $\tn\geq 4$, as we will assume
from now on. We choose $A_{\mu}^{2}$, $A_{\mu}^{3}$,
$A_{\mu}^{4}$ as the $SU(2)_{G}$ gauge fields. Inspection of $N$ above shows
that this group rotates $\xi^{2}$, $\xi^{3}$, $\xi^{4}$ into each other, but
leaves the other $\xi^{\ti}$ unchanged. Thus, no tensor fields have to be
introduced.
The resulting scalar potential (\ref{PR2}) turns out to be
\begin{equation}
P^{(R)}=-\frac{1}{2}(\varphi^{1})^{2}\left[ (\varphi^{2})^{2}+
(\varphi^{3})^{2}+(\varphi^{4})^{2}\right] +
\frac{3}{2}\frac{1}{\varphi^{1}}\,\, ,
\end{equation}
which again does not admit any ground states for the physically interesting
region $\varphi^{1}>0$.

Similar conclusions hold true when one introduces tensor fields by gauging an
 additional $SO(2)$, which is possible when $\tn>5$. Just as in the generic
Jordan case,
  this $SO(2)$ can be chosen to rotate $\xi^{5}$ and $\xi^{6}$ into each
other and thus
  requires the dualization of $A_{\mu}^{5}$ and $A_{\mu}^{6}$
to tensor fields. The total scalar potential then gets an additional
contribution, $P$,
 which turns out to be
\begin{displaymath}
P=\frac{1}{8}(\varphi^{1})^{5}\left[ (\varphi^{5})^{2}+
(\varphi^{6})^{2}\right].
\end{displaymath}
Again, it is easy to see that the combined potential $P_{tot}=P+P^{(R)}$
does not have any critical points for  $\varphi^{1}>0$.

To conclude, at least when the scalar manifold $\mathcal{M}$ is a symmetric
space,
 the $SU(2)_{R}$ gauging leads to a total scalar potential which does not
have any
  critical points.

One also notes that the gauge coupling $g_{R}$ for $SU(2)_{R}$
is related to $g$ (\ref{grg}), which is, of course, a consequence of the
fact that we are gauging a diagonal subgroup of $SU(2)_{R}\times SU(2)_{G}$.
This implies that one cannot tune the relative coupling constants as in
the gaugings of $U(1)_{R}\times K$ in order to change the properties of
critical
 points of the scalar potential $P_{tot}=P+P^{(R)}$ \cite{gz99,gz2} (if such
 critical points were to exist for some of the models we have not studied in
this
 paper).
Hence, $SU(2)_{R}$-gauged supergravity theories are much more rigid than
their $U(1)_{R}$-gauged relatives.

\textbf{Note added}:
 This paper appeared contemporaneously with reference \cite{Cer}
 on the general $\mathcal{N}=2$
$d=5$ supergravity including hypermultiplets and $SU(2)_{R}$
gauging.
 Where these two papers overlap, the authors of \cite{Cer} have found
the results of their revised version to be consistent with our
results.

{\bf Acknowledgements:} We would like to thank Eric Bergshoeff, Renata Kallosh,
Andrei Linde and Toine van Proeyen for fruitful discussions.

\end{document}